# GEO satellites on-orbit repairing mission planning with mission deadline constraint using a large neighborhood search-genetic algorithm

Peng Han[a], Yanning Guo[a], Chuanjiang Li[a*], Hui Zhi[a], Yueyong Lv[a]

a. School of Astronautics, Harbin Institute of Technology, Harbin 150001, China

**Abstract**: This paper proposed a novel large neighborhood search-adaptive genetic algorithm (LNS-AGA) for many-to-many on-orbit repairing mission planning of geosynchronous orbit (GEO) satellites with mission deadline constraint. In the many-to-many on-orbit repairing scenario, several servicing spacecrafts and target satellites are located in GEO orbits which have different inclination, RAAN and true anomaly. Each servicing spacecraft need to rendezvous with target satellites to perform repairing missions under limited fuel. The mission objective is to find the optimal servicing sequence and orbit rendezvous time of every servicing spacecraft to minimize total $\Delta v$ cost of all servicing spacecrafts with all target satellites repaired. Firstly, a time-dependent orbital rendezvous strategy is proposed, which can handle the mission deadline constraint. Besides, it is also cost-effective compared with the existing strategy. Based on this strategy, the many-to-many on-orbit repairing mission planning model can be simplified to an integer programming problem, which is established based on the vehicle routing problem with time windows (VRPTW) model. In order to efficiently find a feasible optimal solution under complicated constraints, a hybrid adaptive genetic algorithm combining the large neighborhood search procedure is designed. The operations of "destroy" and "repair" are used on the elite individuals in each generation of the genetic algorithm to enhance local search capabilities. Finally, the simulations under different scenarios are carried out to verify the effectiveness of the presented algorithm and orbital rendezvous strategy, which performs better than the traditional genetic algorithm.

---

[*] Corresponding author
  Email address: lichuan@hit.edu.cn

# 1 Introduction

With the rapid development of aerospace technology, on-orbit servicing (OOS) compromising refueling, repair, assembly, debris removal has attracted much attention from researchers and engineers[1-2]. During the past decades, one of the most successful OOS missions is the upgrade and repair of the Hubble Space Telescope achieved by space shuttles[3-4]. However, the space shuttle was permanently retired in 2011 due to its high cost. In the future, with the increment in space exploration, the cost of on-orbit servicing needs to be further reduced. Thus, the OOS modes of one-to-many and many-to-many is proposed, i.e., one or more servicing spacecraft (SSc) provide services to multiple satellites successively. How to effectively manage and schedule multiple servicing spacecraft to obtain as many service benefits as possible needs to be solved urgently. Therefore, the study of mission planning and orbit maneuver strategy of multiple servicing spacecraft has great significance.

This paper mainly focuses on the repairing of geosynchronous satellites. It is known that nearly hundreds of high-value satellites are parked in GEO which play an important role in earth observation, communication relay, and meteorological observation. Once the satellite is broken or run out of fuel during daily operation, building and launching a new satellite is more costly than repair the old one. Besides, the geosynchronous orbit is a precious space resource, the broken satellite will cause a waste of GEO resources[5]. Therefore, building an OOS system for GEO satellites is necessary and cost-effective.

Many researchers have studied the OOS mission planning problem for multiple satellites in the past few years, which can be divided into three major categories based on the OOS mission type. (1) On-orbit refueling problem[6-10] which is necessary to consider how to reasonably allocate the servicing spacecraft's limited fuel to orbit maneuver or target satellites. (2) Debris removal problem[11-14]. For the debris removal mission, the main difference from other OOS missions is that the SSc needs to take the target to the grave orbit after rendezvous to the target. (3) Mixed types[15-16]. In the mixed OOS mission scenario, one obvious flaw is that every SSc needs to carry a variety of loads to meet different OOS missions, so that the SSc will have a huge dry weight which leads to more fuel consumption during the orbital maneuver. To authors' best knowledge, the research on on-orbit repairing mission planning is still rare. A typical example of on-orbit repairing is replacing a failed battery with a new battery and assisting a solar array that failed to deploy properly[2]. The lack of power supply may bring about permanent damage to some equipment which causes enormous economic loss to ground users. Therefore, compared with the other OOS mission, the on-orbit repairing mission usually has more urgent mission completion time. Thus, we mainly focus on the many-to-many on-orbit repairing mission planning problem with mission deadline constraints for GEO satellites which is a combinatorial-continuous problem for simultaneously determining the servicing route and orbital rendezvous time of every SSc.

Different types of many-to-many OOS missions can be seen as a multi-spacecraft continuous rendezvous problem[17]. To solve this problem, the orbital rendezvous strategy must be clarified to simulate the SSc maneuver from one target satellite to another. In this paper, the SSc and target satellites are located in the GEO with different inclination, RAAN and true anomaly. Existing literatures usually use a time-independent orbital rendezvous model in which the process of phasing maneuver is usually ignored[8-10, 18]. However, the orbital rendezvous time must be controlled and determined to deal with the mission deadline constraint. Some literatures also propose the time-relevant orbit maneuver strategy: Ref.[15] uses

Lambert algorithm to determine the rendezvous trajectory. Based on the Lambert algorithm, Lorenzo[17] proposes a four-impulse strategy for multirendezvous trajectories optimization problem. However, it is only applicable to coplanar orbital transfer. Inspired by Ref.[10], a mixed time-dependent orbital rendezvous strategy combing the planar change and phasing maneuver is presented in this paper. Simultaneously, the mixed strategy can usually generate a more cost-effective trajectory than the Lambert algorithm. The orbit rendezvous time can be specified by determining the number of revolutions for the phasing maneuver, hence the mission planning problem can be transformed from the combinatorial-continuous problem to an integer programming problem, moreover, the solution space can be further reduced. Thus, the mission planning model is established based on the VRPTW [19] model, in which the servicing route and number of revolutions for every SSc needs to be determined.

To solve the many-to-many OOS mission planning problem which is obviously NP-hard, some researchers used the deterministic algorithm, e.g. column-generation technique[12], branch and bound[20]. Those method usually needs to approximate the mathematical model to a linear model. Besides, those algorithms are very time-consuming even ineffective for large-scale problems. Intelligent metaheuristic algorithms are mostly used in the existing literature like particle swarm algorithm[8, 10, 15], ant colony algorithm[9, 13] etc. Intelligent algorithms usually have a flexible form that can handle complex constraints and objective functions. Genetic algorithm (GA) is also a metaheuristic algorithm which is widely used in on-orbit refueling problems[7], deep-space observation[21] and earth-observing scheduling[22]. GA has a great ability of global exploration in the solution space, however, it does not employ a local search and is easy to fall into local optima[23]. Therefore, it may degrade the solution quality even get an unfeasible solution under tight mission time constraints. To make up for this defect, we present a hybrid algorithm called large neighborhood search adaptive genetic algorithm (LNS-AGA). The large neighborhood search (LNS) [24-26] that has been widely used in routing and scheduling problems is adopted to enhance the general GA's local search capabilities. In addition, the adaptive crossover probability and mutation probability are used in the GA procedure to prevent falling into local optima and speed up the iterative efficiency.

The main contribution of this paper can be summarized as follows:

(1) The mission deadline constraint is considered in the on-orbit repairing mission planning problem, which has a significant meaning in practice.

(2) A time-dependent mixed orbital rendezvous strategy is proposed to simulate the SSc maneuver from one target satellite to another, which is more cost-effective than the Lambert algorithm used in [15]. Then, the mission planning problem can be simplified to an integer programming problem, the solution space can be further reduced.

(3) A hybrid LNS-AGA is designed for solving the mission planning problem, which performs a local search procedure that can have better performance than original genetic algorithm.

The rest of the paper is organized as follows. In Section 2, the mission planning problem is further clarified including a more specific description of the mission scenario, design of the orbital rendezvous strategy and the mathematical model's establishment. After that, Section 3 introduces the design of hybrid LNS-AGA algorithm used for effectively solving this problem. Later on, Section 4 presents one practical scenario and several comparative experiments to demonstrate the effectiveness of the proposed methods. Finally, Section 5 gives some conclusions of this paper and the future path of this research topic.

## 2  Problem formulation

### 2.1  Many-to-many on-orbit repairing mission scenario

In the many-to-many on-orbit repairing scenario, multiple repairing required target satellites and servicing spacecrafts (SSc) are running on GEO, with different orbital inclinations, phase angles, and right ascension of ascending node (RAAN) angles. Every target satellite needs to be repaired by exactly one SSc and every SSc can provide on-orbit repairing for any target satellite. The on-orbit repairing mission must be completed before the pre-set mission deadline. Before performing on-orbit repairing to a target, the SSc should first perform orbital rendezvous to close to the target and dock with the target. The SSc will make a velocity increment by impulse during preform orbital rendezvous. Every SSc could carry a limited fuel for orbital rendezvous, so the SSc must satisfy the maximum velocity increment constraint throughout the whole mission process. Otherwise, it will not be able to complete the scheduled on-orbit repairing mission. After completing all on-orbit service tasks, in order not to affect the regular operation of the repaired targets, SSc needs to perform orbital maneuver, away from the last target slowly, and arrive at a parking orbit waiting for the next mission. The cost of orbit maneuver of this process will be ignored. The optimal on-orbit repairing strategy needs to determine the repairing route of every SSc, and the orbital maneuver strategy of every on-orbit repairing process, the velocity increment cost of all SSc is minimized.

Fig. 1 gives a feasible solution to an instance of a many-to-many on-orbit scenario with 6 target satellites and 2 SSc. In this solution, the two SSc will start from different initial orbit and dock with 3 target satellites in order. The SSc1 will perform an on-orbit repairing for target satellites 6, 3 and 4 in order. The SSc1 will perform an on-orbit repairing for target satellites 1, 2 and 5 in order. After completing the repairing mission, those two SSc will go to the parking orbit slowly. The mission deadline constraint and maximum velocity increment constraint mentioned above will be satisfied during the whole procedure. Consequently, all target satellites are repaired exactly once by SSc.

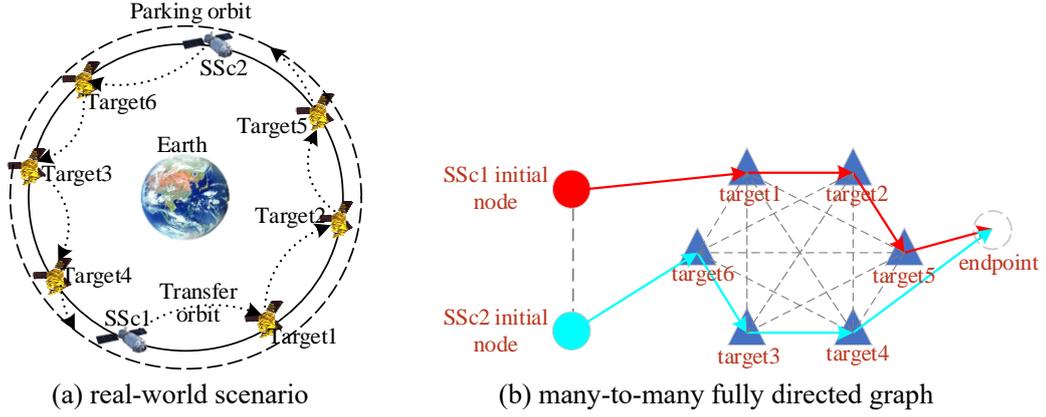

(a) real-world scenario  (b) many-to-many fully directed graph

Fig. 1 many-to-many on-orbit repairing scenario

**2.2 Time-dependent orbital rendezvous strategy**

The SSc needs a long-distance orbital maneuver to reach every target, then the SSc can dock with the target and perform an on-orbit repairing mission. In this paper, only the $\Delta v$ cost of the long-distance orbital maneuver process is considered, and the $\Delta v$ cost of the docking process is ignored.

The SSc need to perform a maneuver procedure to let the SSc and target have the same inclination, RAAN and true anomaly to perform the docking process. This procedure can be achieved by planar maneuver and phasing maneuver. Besides, according to Ref. [27-28], the perturbation effect can be ignored on GEO within one year and the repairing mission usually needs to be completed in a short time. Therefore, the two-body orbital dynamics will be used to generate the trajectory of SSc and target satellite.

**2.2.1 Planar change**

Planar change lets the SSc run on the same orbital plane with the target through one impulse maneuver. As shown in Fig. 2,SSc can only maneuver at the intersection of two orbital planes. Therefore, SSc has two chances of plane change maneuver in one orbital period.

The dihedral angle $\alpha$ between two orbital planes is given by

$$\cos\alpha = \sin I_s \sin I_t \cos(\Omega_s - \Omega_t) + \cos I_s \cos I_t \tag{1}$$

where $I_s, I_t, \Omega_s, \Omega_t$ represent the inclination and RAAN of the SSc and target, respectively. the intersection points of the two orbital planes can be calculated by

$$\begin{cases} \boldsymbol{r}_{m1} = \boldsymbol{h}_s \times \boldsymbol{h}_t \\ \boldsymbol{r}_{m2} = -\boldsymbol{r}_{m1} \end{cases} \tag{2}$$

where $\boldsymbol{h}_s, \boldsymbol{h}_t$ represent the angular momentum of SSc and target, respectively. the angular momentum of the orbit $\boldsymbol{h}$ can be calculated by

$$\boldsymbol{h} = \sqrt{\frac{a}{\mu}} \begin{bmatrix} \cos\Omega & -\sin\Omega & 0 \\ \sin\Omega & \cos\Omega & 0 \\ 0 & 0 & 1 \end{bmatrix} \begin{bmatrix} 1 & 0 & 0 \\ 0 & \cos i & -\sin i \\ 0 & \sin i & \cos i \end{bmatrix} \begin{bmatrix} 0 \\ 0 \\ 1 \end{bmatrix} \tag{3}$$

where $a$ is the orbit radius of GEO orbit, $\mu$ is the Earth's gravitational constant, $\boldsymbol{h}$ is constant without considering any perturbation.

Since SSc and target are located on the circular GEO orbit, it is also necessary to ensure that the velocity remains unchanged after the impulse maneuver, only the velocity direction will be changed. We define $\Delta \boldsymbol{v}_1$ as an orbital plane change impulse maneuver of $\alpha$, it can be calculated by:

$$\begin{aligned} \Delta \boldsymbol{v}_1 &= \boldsymbol{v}_t - \boldsymbol{v}_s \\ &= 2\|\boldsymbol{v}_t\|\sin(\Delta i/2) \end{aligned} \tag{4}$$

Where $\boldsymbol{v}_s$, $\boldsymbol{v}_t$ is the velocity of SSc and target at the maneuver point, respectively.

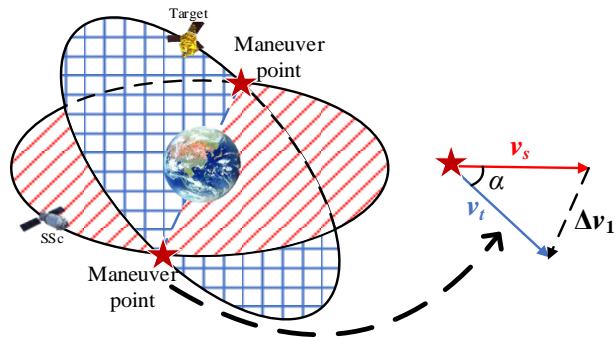

Fig. 2 Sketch of orbital plane change maneuver

After the last repairing mission is finished, the SSc needs to coast for a while on the original orbit until it reaches the nearest plane change maneuver point for the next target. Given a SSc initial state of $\{r_s^0, v_s^0\}$, the coast time before planar change maneuver can be calculated by

$$t_{coast} = \begin{cases} \arccos\left(\dfrac{r_{m1} \cdot r_s^0}{\|r_{m1}\|\|r_0\|}\right) T_{GEO} \Big/ 2\pi, & (r_s^0 \times r_{m1}) \cdot h_s > 0 \\ \left(\pi - \arccos\left(\dfrac{r_{m1} \cdot r_s^0}{\|r_{m1}\|\|r_s^0\|}\right)\right) T_{GEO} \Big/ 2\pi, & (r_s^0 \times r_{m1}) \cdot h_s < 0 \end{cases} \tag{5}$$

where $T_{GEO}$ is the GEO orbital period.

### 2.2.2 Phasing maneuver

After the plane change maneuver, the SSc and the target will run in the same circular GEO orbit, so they have a fixed phase difference. We define the phasing angle between the SSc and target as

$$\theta = \begin{cases} |(\Omega_s + \omega_s) - (\Omega_t + \omega_t)|, & |(\Omega_s + \omega_s) - (\Omega_t + \omega_t)| \leq \pi \\ 2\pi - |(\Omega_s + \omega_s) - (\Omega_t + \omega_t)|, & |(\Omega_s + \omega_s) - (\Omega_t + \omega_t)| > \pi \end{cases} \tag{6}$$

where $\omega_s$ and $\omega_t$ denote the true anomaly at the same time of SSc and target, respectively. As shown in Fig. 3, SSc needs to perform an impulse maneuver and then enter the phasing orbit. The phasing orbit is tangent to the original orbit at the maneuver point. After the SSc runs a certain number of turns in the phasing orbit, it returns to the maneuver point. It should be ensured that the target also runs exactly to the tangent point, then the SSc performs an impulse maneuver to meet the target. We know that the SSc can only meet the target at a specific time within one orbital period. So, the phasing time is

$$t_{phase} = \frac{2\pi k_{GEO} + \theta}{2\pi} T_{GEO} \tag{7}$$

where $k_{GEO}$ is target's number of revolutions in phasing maneuver, $T_{GEO}$ is the GEO orbital period. The semimajor axis of the phasing orbit can be calculated by

$$a_{phase} = r_{GEO} \left(\frac{2\pi k_{GEO} + \theta}{2\pi k_{phase}}\right)^{2/3} \tag{8}$$

where $r_{GEO}$ is the orbit radius of GEO satellite, $k_{phase}$ denotes target's number of revolutions in phasing maneuver. Then the $\Delta v$ for phasing maneuver can be calculated by

$$\Delta v_{phase} = 2\sqrt{\mu} \left| \sqrt{\frac{2}{r_{GEO}} - \frac{1}{a_{phase}}} - \sqrt{\frac{1}{r_{GEO}}} \right| \tag{9}$$

Inspection of Eq.(8) and Eq.(9) reveals that when $k_{GEO}$ is determined, $k_{phase}=k_{GEO}$ must be satisfied in order to have the minimum $\Delta v_{phase}$ cost.

Also, we can know that these two impulse maneuvers ($\Delta v_2$ and $\Delta v_3$) for phasing maneuver is in line with the target velocity direction at the maneuver point. Based on Fig. 3, we could know that

$$\begin{cases} \Delta \boldsymbol{v}_2 = -\frac{1}{2} \Delta v_{phase} \, \text{sgn}(\theta) \frac{\boldsymbol{v}_m}{\|\boldsymbol{v}_m\|} \\ \Delta \boldsymbol{v}_3 = -\Delta \boldsymbol{v}_2 \end{cases} \tag{10}$$

where $\boldsymbol{v}_m$ is the target velocity at the maneuver point, it can be calculated by the orbital elements of the target.

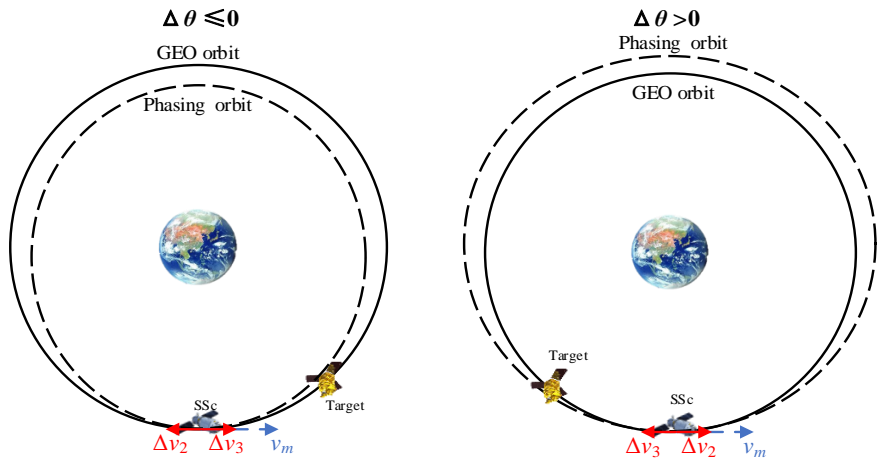

Fig. 3 Sketch of the phasing maneuver

### 2.2.3 Mixture of planar change and phasing maneuver

Each SSc and target will keep a constant phase difference because they run in the circular GEO orbit. So, the SSc can perform the phasing maneuver at any time, hence we can set the phasing maneuver point at the plane change point. As shown in Fig. 4, the plane change maneuver impulse $\Delta v_1$ and the first phasing maneuver impulse $\Delta v_2$ can be combined into one impulse maneuver. Based on that, the total $\Delta v$ cost for one orbital rendezvous can be expressed as

$$\begin{aligned} \Delta v &= \|\Delta \boldsymbol{v}_1 + \Delta \boldsymbol{v}_2\| + \|\Delta \boldsymbol{v}_3\| \\ &= \sqrt{\|\Delta \boldsymbol{v}_1\|^2 + \|\Delta \boldsymbol{v}_2\|^2 + 2\|\Delta \boldsymbol{v}_1\|\|\Delta \boldsymbol{v}_2\|\cos\alpha} + \|\Delta \boldsymbol{v}_2\| \end{aligned} \quad (11)$$

$\alpha$ denotes the angle between $\Delta \boldsymbol{v}_1$ and $\Delta \boldsymbol{v}_2$. It is obvious that after the combination of the two strategies, there will be less $\Delta v$ cost than three impulse maneuver methods. Given a certain phasing revolution number $k_{GEO}$, the total time of rendezvous maneuver can be calculated by

$$t_r = t_{coast} + t_{phase} \quad (12)$$

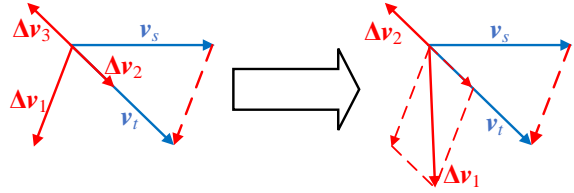

Fig. 4 Sketch of mixture of the two maneuver strategies

## 2.3 Mathematical model for the mission planning problem

In general, the mission planning problem is formulated as a combinatorial-continuous problem when selecting the servicing route and the orbital rendezvous time of every SSc as the decision variable. Based on the orbital rendezvous described in Section 2.2, the rendezvous time can be determined by the number of revolutions of phasing maneuver for every rendezvous, which is an integer variable. By selecting the integer variable as the decision variable, the problem can be transformed to an integer programming problem. Therefore, the many-to-many on-orbit repairing mission planning model can be established based on the modeling method of widely studied VRPTW[19] problem. As shown in Fig. 1 (a), all SSc and targets can be defined on a fully directed graph $G = (N, A)$, where $N$ is a set of nodes comprises a subset $N_s$ of $m$ SSc initial nodes, a subset $N_t$ of $n$ target nodes and a virtual endpoint $e$. Setting a virtual endpoint can help us easily model and construct constraints. $A$ is the set of edges connected by any two nodes in $N$.

A fleet of SSc will be represented by $V$ consists of several identical SSc with a total maximum velocity increment. Each SSc starts from its corresponding initial node, reaches the target node, and finally reaches the virtual endpoint after completing the on-orbit repairing mission. Define a decision variable $x_{ijk}$ to describe whether each SSC passes through an edge in edge set $A$, $x_{ijk}=1$ means that the edge $(i,j)$ is traversed from node $i$ to $j$ by the SSc $k$. the SSc cannot perform an orbital maneuver between two SSc initial nodes because that is meaningless. Each SSc initial node can only send one SSc, which has the same number as the initial node. All SSc should arrive at the virtual endpoint $e$ after completing the on-orbit repairing mission. The transfer cost between nodes is related to orbital maneuver time and orbital position. The orbital parameters of each node are fixed, so the orbital maneuver time between nodes will be taken as the decision variable. In this paper, based on the orbital rendezvous strategy described in Section 2.2, it will be replaced by the target's number of revolutions in phasing maneuver. Each target node should be visited and repaired exactly once by a single SSc within the mission deadline. The objective is to determine the optimal repairing sequence and orbital maneuver time of each SSc while minimizing the total $\Delta v$ cost of all SSc.

The remaining parameters which will be used in the mission planning model are defined in Table 1.

Table 1 Partial parameters used in mission planning models

| Parameter | Definition |
| --- | --- |
| $s_{ik}$ | Arrival time of $k$-th SSc to target node $i$, $i \in N_t$, $k \in V$ |
| $t_r^{ij}$ | Orbital maneuver time between node $i$ and $j$, $i, j \in N$ |
| $c_{ij}$ | $\Delta v$ cost from node $i$ to $j$, $i, j \in N$ |
| $t_{di}$ | On-orbit repairing time of target $i$, $i \in N$ |
| $T_{end}$ | Mission deadline time |
| $\Delta v_m$ | Maximum velocity increment of every SSc |

| | | |
|---|---|---|
| | $n_{ij}$ | Number of revolutions in phasing maneuver |
| $E_i = \{a_i, e_i, i_i, \Omega_i, \eta_i, \omega_i\}$ | | Orbital elements of node $i$ |

Based on the definition, the mission planning model can be formulated as follows

(1) Decision variables: $x_{ijk}$, $n_{ij}$, $\forall i \in N, j \in N, k \in V$.

(2) Objective function:

$$\min \sum_{k \in V} \sum_{i \in N} \sum_{j \in N} c_{ij} x_{ijk} \tag{13}$$

(3) Constraints:

$$\sum_{k \in V} \sum_{j \in N} x_{ijk} = 1, \forall i \in N_t \tag{14}$$

$$\sum_{i \in N} \sum_{j \in N_s} x_{ijk} = 0, \forall k \in V \tag{15}$$

$$\sum_{j \in N} x_{ijk} = 1, \forall k \in V, i \in N_s, k = i \tag{16}$$

$$\sum_{j \in N} x_{ijk} = 0, \forall k \in V, i \in N_s, k \neq i \tag{17}$$

$$\sum_{i \in N} x_{ihk} - \sum_{j \in N} x_{hjk} = 0, \forall k \in V, h \in N_t \tag{18}$$

$$\sum_{i \in N} x_{i,e,k} = 1, \forall k \in V \tag{19}$$

$$c_{ie} = 0, \forall i \in N \tag{20}$$

$$s_{ik} + t_{di} + t_r^{ij} - M(1 - x_{ijk}) \leq s_{jk}, \forall i \in N, j \in N, k \in V \tag{21}$$

$$0 < s_{ik} + t_{di} \leq T_{end}, \forall i \in N, k \in V \tag{22}$$

$$[t_{ij}, c_{ij}] = f_{OM}(E_i, E_j, s_{ik}, n_{ij}), \forall i \in N_s \cup N_t, j \in N_s \cup N_t \tag{23}$$

$$\sum_{i,j \in N} c_{ij} x_{ijk} \leq \Delta v_m, \forall k \in V \tag{24}$$

$$x_{ijk} \in \{0,1\}, \forall i \in N, j \in N, k \in V \tag{25}$$

$$1 \leq n_{ij} < T_{end} / T_{GEO}, n_{ij} \in \mathbb{Z}, \forall i \in N, j \in N \tag{26}$$

The objective function (13) is to minimize the total $\Delta v$ cost of all SSc. Constraint (14) ensures that one target can be repaired only once by exactly one SSc. Constraint (15) ensures the SSc can only start from the SSc initial nodes and cannot maneuver to any SSc initial nodes. Constraint (16) and (17) means that each SSc initial node can only send exactly one SSc with the same number as the initial node. Constraint (18) guarantees the continuity of each repairing route. Constraint (19) ensures that each SSc repairing route will terminate at the virtual endpoint. Constraint (20) indicates that the $\Delta v$ cost of the edges connecting the endpoint are 0. Constraint (21) is a time constraint, it states that a SSc cannot arrive at node $j$ before $s_{ik} + t_{di} + t_{ij}$ if it travels from target $i$ to $j$. Constraint (22) guarantees that every repair task will end before the mission deadline. Constraint (23) indicates that the orbital maneuver time and $\Delta v$ cost of any edges in $A$ except those connecting virtual endpoint will be calculated by the nonlinear function $f_{OM}$, while its calculation method is described in section 2.2. The maximum velocity increment constraints of every SSc are given in (24). Finally, constraint (25) and (26) give the restriction on the decision variables.

## 3  Optimization method

### 3.1  Design of adaptive genetic algorithm

#### 3.1.1  Encoding mechanism

In the genetic algorithm, the conversion method that converts the feasible solution of the problem from its solution space to the search space that the genetic algorithm can handle is called encoding[29-30]. The design of encoding rules plays a vital role in the solving efficiency of genetic algorithm. The sequence coding method is used in this paper for encoding the decision variable $x_{ijk}$. Assume there is an on-orbit repairing mission scenario consist of $m$ targets and $n$ SSc, to solve the mission planning by genetic algorithm, the chromosome length will first be fixed length to $m+n-1$. Every chromosome will consist of a random sequence from 1 to $m+n-1$, so every gene in the chromosome cannot be repeated and it represents the on-orbit repairing route of all SSc. Those genes numbered 1 to $m$ represent the corresponding target number. The genes numbered $m+1$ to $m+n-1$ are split genes, which divide the entire target sequence into *n* fragments, and the sequence of each fragment corresponds to the on-orbit repairing route of the corresponding SSc. Based on the encoding method, each chromosome will uniquely correspond to a feasible solution of $x_{ijk}$, but a feasible solution may correspond to multiple coding chromosomes. This encoding method can assure the feasible solution of $x_{ijk}$ will always satisfy the route constraints (14)-(19).

Fig. 5 give a feasible example of the coding mechanism. This example consists of 3 SSc and 8 targets, the SSc initial node number is 1 to 3, the target node number is 4 to 11. So, a chromosome of length 10 will be obtained through the encoding mechanism. Since the coded chromosome does not obviously contain the SSc initial node number, the target number in the chromosome will start from 1 to 8 with two split genes numbered 9 and 10. The chromosome is divided into 3 parts by the split genes: **5-3-1**, **2-7-6**, **4-8**. It means that SSc1 will start from SSc initial node **1** and go to the target node **8**, **6**, **4** in order to perform an on-orbit repairing mission; SSc2 will start from SSc initial node **2** and go to the target node **5**, **10**, **9** successively; SSc3 will start from SSc initial node **3** and go to the target node **7**, **11** in order. After complete the mission, all SSc will go to the parking orbit.

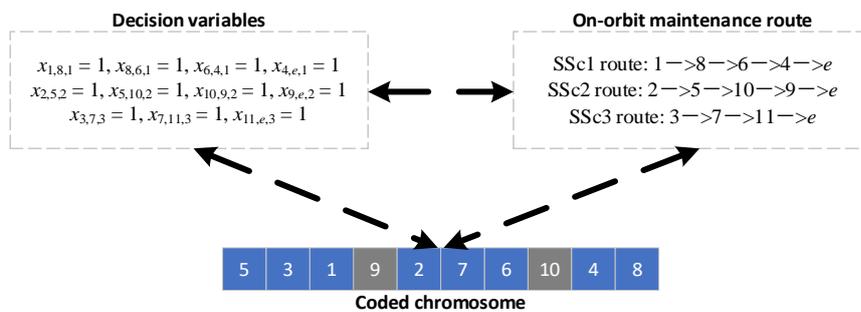

Fig. 5 Examples of coding mechanism

### 3.1.2 Fitness function

The fitness function is used to measure the degree of conformity of the generated solution, including the objective function value and the satisfaction of constraints. As we mentioned before, the feasible solution of decision variable generated by the sequence coding method will always satisfy the route constraints (14)

-(19) and value constraint (25)，Therefore, there is no need to measure those constraints in the fitness function. Constraint (20) is an invalid constraint in the process of generating a feasible solution. Only constraints (21)-(24) need to be measured by the fitness function. In this paper, the penalty function method is used to measure the satisfaction of constraints. Thus, the fitness function is given by

$$F = \sum_{k \in V} \sum_{i \in N} \sum_{j \in N} c_{ij} x_{ijk} + \varphi p_1(s_{ik}) + \gamma p_2(x_{ijk}, c_{ij}) \tag{27}$$

The fitness function consists of three parts: the total $\Delta v$ cost of all SSc, that is, Eq.(13); the penalty function of constraint (22) with a constant coefficient $\varphi$; the penalty function of constraint (24) with a constant coefficient $\gamma$. Those two penalty functions $p_1$, $p_2$ are defined as follows:

$$p_1(s_{ik}) = \sum_{k \in V} \sum_{i \in N_t} \max\{(s_{ik} + t_{di} - T_{end}), 0\} \tag{28}$$

$$p_2(x_{ijk}, n_{ij}) = \sum_{k \in V} \max\left\{\left(\sum_{i \in N} \sum_{j \in N} c_{ij} x_{ijk} - \Delta v_m\right), 0\right\} \tag{29}$$

Eq.(28) represents the sum of the duration of every mission completion time beyond the mission deadline. Eq. (29) represents the total $\Delta v$ cost of each SSc that exceeds $\Delta v_m$. Before calculating the fitness function described in Eq. (27)-(29), intermediate variables $s_{ik}$ and $c_{ij}$ need to be calculated by Eq.(21), Eq.(23) and the feasible solution decoded from the chromosome. Notice that before calculating the intermediate variables, another decision variable $n_{ij}$ also need to be given. The method of determine the decision variable $n_{ij}$ will be described in the next section.

### 3.1.3 Genetic operator

(1) selection operator

Elite strategy is first used in the selection operator, which selects the chromosome with the highest fitness function to participate in the next genetic operator. Then, the roulette algorithm is used to select other chromosomes from the remaining population. In the roulette algorithm, the cumulative probability is used to judge whether the chromosome is selected, which is given by

$$F_i = \frac{F(i)}{\sum_{j=1}^{n} F(j)}, \quad i=1,2,...,n \tag{30}$$

$$Q_i = \sum_{j=1}^{i} P_i \tag{31}$$

where $P_i$ is the proportion of fitness value of $i$-th chromosome in the population, $n$ is the number of chromosomes in the remain population. $Q_i$ is the cumulative selected probability from the first chromosome to $i$-th chromosome. The $i$-th chromosome is selected for the next genetic operator when the generated random number $r \in (0,1)$ satisfies $Q_{i-1} < r < Q_i$.

(2) crossover operator

In crossover operator, the adaptive crossover probability, which can enhance the iterative efficiency, is used to determine whether the selected two chromosomes perform the crossover operation:

$$P_c = \begin{cases} P_{c1} - \frac{(P_{c1} - P_{c2})(F_i' - F_{avg})}{F_{max} - F_{avg}}, & F_i' \geq F_{avg} \\ P_{c1}, & F_i' < F_{avg} \end{cases} \tag{32}$$

where $F_i'$ is the larger fitness value of the selected two chromosomes; $F_{max}$ and $F_{avg}$ denote the maximum fitness value and average fitness value of the population, respectively; $P_{c1}$ and $P_{c2}$ is the upper and lower bound of $P_c$.

Partial-Mapped Crossover[31] is used in the crossover operator. The method exchanges the genes between the two randomly chosen positions in the two selected chromosomes. After that, same genes will be existed in one chromosome, which is not permitted based on the encoding method. Three steps are performed to eliminate this phenomenon: (1) For the exchanged two chromosomes, find the genes in the unexchanged part which hold the same value as the exchanged parts and extract them. (2) Match the genes extracted from different chromosome in turn. (3) Exchange the matched genes one by one and put them into the original location. The whole procedure is presented in Fig. 6.

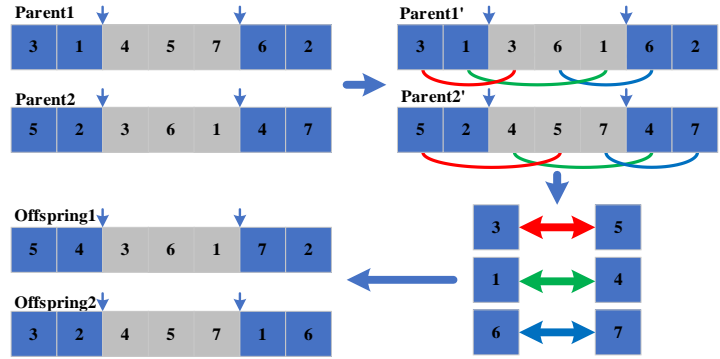

Fig. 6 Crossover operation

(3) mutation operator

Adaptive mutation probability $P_m$ is used to determine whether the selected chromosome is mutated or not. The adaptability of mutation probability can prevent falling into local optima[22]. $P_m$ is given by

$$P_m = \begin{cases} P_{m1} - \dfrac{(P_{m1} - P_{m2})(F_{\max} - F_i)}{F_{\max} - F_{avg}}, & F_i \geq F_{avg} \\ P_{m1}, & F_i < F_{avg} \end{cases} \qquad (33)$$

where $P_{m1}$, $P_{m2}$ is the upper bound and lower bound of $P_m$.

As shown in Fig. 7, for the mutated chromosome, two gene sites are randomly selected and their genes are exchanged.

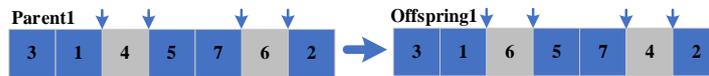

Fig. 7 Mutation operation

## 3.2 Heuristics for determining the decision variable $n_{ij}$

As we can know from Section 2.2, if the repair route of the SSc is determined, the $\Delta v$ cost for planar change is constant, but the $\Delta v$ cost for phasing maneuver is time dependent and is also determined by $n_{ij}$ in this paper. The orbit information of every target is known and it does not change during the mission period. So, the $k$-th SSc's every maneuver point for orbital maneuver is fixed in space. Since the required repair time $t_{di}$ for every target is also fixed, the SSc's every coast time for planar change can be can be calculated by Eq.(5). Therefore, the total phasing maneuver time for $k$ -th SSc $T_{phase}^k$ is given by

$$T_{phase}^k = \sum t_{phase}^{ij} = T_{end} - \sum t_{dj} - \sum t_{coast}^{ij} \tag{34}$$

where $i$, $j$ satisfies $\forall i, j \in N, x_{ijk} = 1$. $t_{phase}^{ij}$ and $t_{coast}^{ij}$ denote the phasing maneuver time and the coast time before orbital plane change maneuver from target $i$ to $j$.

Once the upper-level algorithm determines the repair route, the determination of $n_{ij}$ for the $k$ -th SSc can be formulated as a nonlinear integer programming problem.

(1) Decision variables: $n_{ij}, \forall i, j \in N, x_{ijk} = 1$.

(2) Objective function:

$$\min: 2\sqrt{\mu} \sum_{\substack{i \in N, j \in N, \\ x_{ijk}=1}} \left| \sqrt{\frac{2}{r_{GEO}} - \frac{1}{r_{GEO}\left[1+(\theta_{ij}/2\pi n_{ij})\right]^{2/3}}} - \sqrt{\frac{1}{r_{GEO}}} \right| \tag{35}$$

(3) Constraints

$$\sum_{\substack{i \in N, j \in N, \\ x_{ijk}=1}} \left[ T_{GEO} \left( n_{ij} + \frac{\theta_{ij}}{2\pi} \right) \right] \leq T_{phase}^k \tag{36}$$

$$1 \leq n_{ij} < T_{phase}^k / T_{GEO}, n_{ij} \in \mathbb{N}, \forall i \in N, j \in N, x_{ijk} = 1 \tag{37}$$

The objective function (35) minimizes the total phasing maneuver $\Delta v$ cost of $k$ -th SSc, which is obtained by substituting (8) into (9). Constraints (36) is the mission deadline constraint, it is a variant of (22). Constraints (37) is the value constraints of $n_{ij}$. Solving this NLIP problem can be complicated and time - consuming. In this paper, a simple and effective heuristic algorithm is proposed to quickly give a near optimal solution. The pseudocode for the heuristic is shown in Algorithm1. The main idea of the algorithm is to first divide the total phasing maneuver time $T_{phase}^k$ equally according to the orbital period $T_{GEO}$ into $m$ pieces, then distribute those time pieces evenly to each phasing maneuver procedure, and finally allocate the extra time to the phasing maneuver procedure with a largest phase angle difference.

**Algorithm1:** Heuristics for determining $n_{ij}$

1. **input** $k$-th SSc repairing target sequence set: $L_k$, phasing angle set: $\Theta_k$
2. initialize decision variable $n_{ij}$ of SSc set: $N_l$
3. calculate $T_{phase}^k$ by Eq.(34)
4. $m = floor\left(T_{phase}^k / T_{GEO}\right)$
5. **for** $n_{ij}$ **in** $N_l$
6.     $n_{ij} = \max\{floor(m / size(L_k)), 1\}$
7. **end for**
8.     calculate SSc's mission end time $t_{end}^k$ based $L_k, N_l$ by Section 2.2
9. **if** $t_{end}^k < T_{end}$
10.     $m_r = floor\left((T_{end} - t_{end}^k) / T_{GEO}\right)$
11.     find target $a$ and $b$ in $L_k$ which have a minimum $\theta_{ab}$ in $\Theta_k$
12.     $n_{ab} = n_{ab} + m_r$
13. **end if**
14. **return** $N_l$

## 3.3 Design of large neighborhood search method

### 3.3.1 LNS framework

The large neighborhood search (LNS) was first proposed by Shaw[25], its main idea is to find a better solution in the search space by iteratively destroying and repairing the feasible solutions. In LNS algorithm, the neighborhood is implicitly defined by the *destroy* and *repair* method[25]. The pseudocode for the LNS algorithm is shown in Algorithm2. There are three variables in the pseudocode. The variable $x^b$ is the best solution of the current iteration, $x$ is the feasible solution and $x^t$ is the temporary feasible solution that could be accepted or discarded based on the acceptance rule. Function *destroy*(·) is used to destruct the feasible solution and return an infeasible solution $x^d$. Function *repair*(·) is the repair method that rebuild the destroyed solution, it usually returns a feasible solution built from $x^d$. More details about the destroy and repair method is proposed in Section 3.3.2. and Section 3.3.3, respectively. $c(x)$ denotes the objective function value of solution $x$. The acceptance criteria $accept(x^t, x)$ usually have many forms such as simulated annealing, hill-climber, threshold accepting, record-to-record travel. In this paper, the hill-climber method will be selected as the acceptance criteria, which only accept an improving solution. The hill-climber is also used in Shaw's original LNS paper.

| | **Algorithm2: Large Neighborhood Search** |
|---|---|
| 1 | **input** a feasible solution $x$ |
| 2 | $x^b = x$; |
| 3 | **repeat** |
| 4 | $\quad x^t = repair(destroy(x))$; |
| 5 | $\quad$ **if** $accept(x^t, x)$ **then** |
| 6 | $\quad\quad x = x^t$; |
| 7 | $\quad$ **end if** |
| 8 | $\quad$ **if** $c(x^t) < c(x^b)$ **then** |
| 9 | $\quad\quad x^b = x^t$; |
| 10 | $\quad$ **end if** |
| 11 | **until** stop criterion is met |
| 12 | **return** $x^b$ |

### 3.3.2 Destroy method

The destroy method is used to remove a fixed number target from the feasible routes. We sequentially remove the targets which are similar to each other. The general idea is to delete some similar targets, because it can make similar targets mixed together, to create a new and perhaps better solution. Suppose the removed targets are very different from each other. In that case, we will not get a better feasible solution when re-inserting the targets, because those removed targets can only be inserted into their original positions or some wrong positions. A relatedness measure $R(i, j)$ is defined to evaluate the similarity of two targets $i$ and $j$. The lower $R(i, j)$ is, the less related are the two targets. $R(i, j)$ is given by

$$R(i, j) = 1 / (C'_{ij} + V_{ij}) \tag{38}$$

where $V_{ij}$ is a parameter that evaluates whether target $i$ and $j$ are repaired by the same SSc, $V_{ij}$ is given by

$$V_{ij} = \begin{cases} 0, \sum_{k \in V} x_{ijk} = 1 \\ 1, \sum_{k \in V} x_{ijk} = 0 \end{cases} \tag{39}$$

$C'_{ij}$ is a normalized parameter that approximately evaluates the cost of getting to target $j$ from $i$. $C'_{ij}$ is defined as follows

$$C'_{ij} = C_{ij} / \max\{C_{mn}\}, \forall m \in N_t, n \in N_t \tag{40}$$

In Shaw's[25] paper, $C_{ij}$ denotes the distance between the two nodes for the classical VRP problem. It approximately represents the transfer cost of the vehicle between two nodes. In this paper, the distance between the two target satellites changes over time and the transfer cost is also time-dependent according to Section 2.2, so we use the orbital difference to approximately evaluate the transfer cost between the two targets. $C_{ij}$ is given by

$$C_{ij} = \beta |\alpha_{ij}| + (1-\beta)\theta_{ij}, 0 < \beta < 1 \tag{41}$$

where $\alpha_{ij}$ is the dihedral angle between targets $i$ and $j$, $\theta_{ij}$ denotes the phase difference between the two targets. $\beta$ is a determinism parameter used to adjust the weight of orbital plane difference and phase difference. Generally, the longer the mission period, the larger the parameter $\beta$ set, because when the maneuver time increase the phasing maneuver $\Delta v$ cost will be decreased, but the $\Delta v$ cost of orbital plane change is time-independent.

The pseudocode for the remove procedure is shown in Algorithm3. The procedure initially chooses a random target to remove. Then it runs into an iteration loop for selecting the targets similar to the latest removal target to remove until the removed target number exceeds the maximum remove rate. Determinism parameter $p \geq 1$ is used to add the randomness of the target removal process[24] (a lower value of $p$ corresponds to more randomness) .

**Algorithm3: Remove Procedure**

```
 1  input a feasible solution x, remove rate q
 2  initialize removal set: x^r = ∅, destroyed set: x^d = x
 3  randomly remove a target t^r from x^d
 4  x^r = x^r ∪ t^r
 5  repeat
 6      calculate R(t_r, t), ∀t ∈ x^d
 7      set: L = sort descending t based on R(t_r, t), ∀t ∈ x^d
 8      choose a random number y from the interval [0,1)
 9      t^r = select the floor(y^p) th target from L
10      x^r = x^r ∪ t^r
11  until size(x^r) ≥ size(x) × q
12  return x^r, x_d
```

### 3.3.3 Repair method

We present the repair method based on the farthest insertion heuristic. The core idea of the algorithm is to first insert the removed targets with the highest "insertion cost". The "insertion cost" of the removed target $i$ is defined as

$$c_i = \min\{\Delta f_{i,k}\}, \forall k \in V \tag{42}$$

where $\Delta f_{i,k}$ denotes the objective value change caused by inserting the removed target $i$ into $k$-th SSc route at the position that increases the objective value the least. If all the constraints cannot be satisfied when inserting target $i$ into $k$-th SSc route, the value of $\Delta f_{i,k}$ is set to $\infty$. As shown in Fig. 8, the "insertion cost" means the minimum objective value increment of inserting target $i$ at its best position overall. We select the target with the largest insertion cost and insert it into the minimum cost position. Repeat the above operations until all the removed targets are inserted into the original broken routes. The pseudocode is shown in algorithm4.

Note that if a greedy strategy is adopted, the target with the lowest insertion cost should be inserted first, but this may cause the target with a larger insertion cost to be difficult to find a feasible insert position in the end of the insert procedure due to the violation of the mission deadline constraint or the SSc velocity increment constraint. That will reduce the efficiency of the re-insertion algorithm. Therefore, we present the farthest insertion heuristic to first insert the target that is more difficult to insert to ensure that the re-insertion algorithm can generate a feasible solution to the greatest extent.

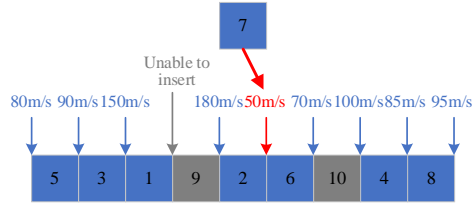

Fig. 8 Explanation of "insertion cost"

| Algorithm4: Re-insert procedure |
|---|
| 1  **input** removal set: $x^r$, destroyed set: $x^d$ |
| 2  initialize repaired set: $x^t = x^d$ |
| 3  **repeat** |
| 4      **for** $t_i^r$ **in** $x^r$ |
| 5          Calculate $\Delta f_{i,k}$ of $t_i^r$ based on $x^t$ |
| 6          Find the minimum cost position $P_{i,k}$ of $t_i^r$ in $x^t$ |
| 7      **end for** |
| 8      find $t_i^r$ in $x^r$ with maximum $\Delta f_{i,k}$ |
| 9      insert $t_i^r$ into $P_{i,k}$ in $x^t$ |
| 10     remove $t_i^r$ from $x^r$ |
| 11 **until** $x^r == \emptyset$ |
| 12 **return** $x^t$ |

## 3.4 Framework of the LNS-AGA

The flowchart of the proposed LNS-AGA algorithm is presented in Fig. 9. We know that GA has a great performance on global search, but it does not employ a local search mechanism. In order to make up for this defect, a LNS procedure is added after the GA operator. In each iteration process of LNS-AGA, after performing the GA operation, the algorithm will select a fixed number of $n$ chromosomes with highest fitness value and input them into the LNS process. Those selected chromosomes will repeatedly perform the remove operator and re-insert operator until the fitness value is improved or the LNS iteration number exceeds the set maximum number of iterations. Based on the combination of GA and LNS, the algorithm will have not only great global search performance but also have strong local optimization ability, this will improve the algorithm's convergence speed to optimal and near optimal solutions.

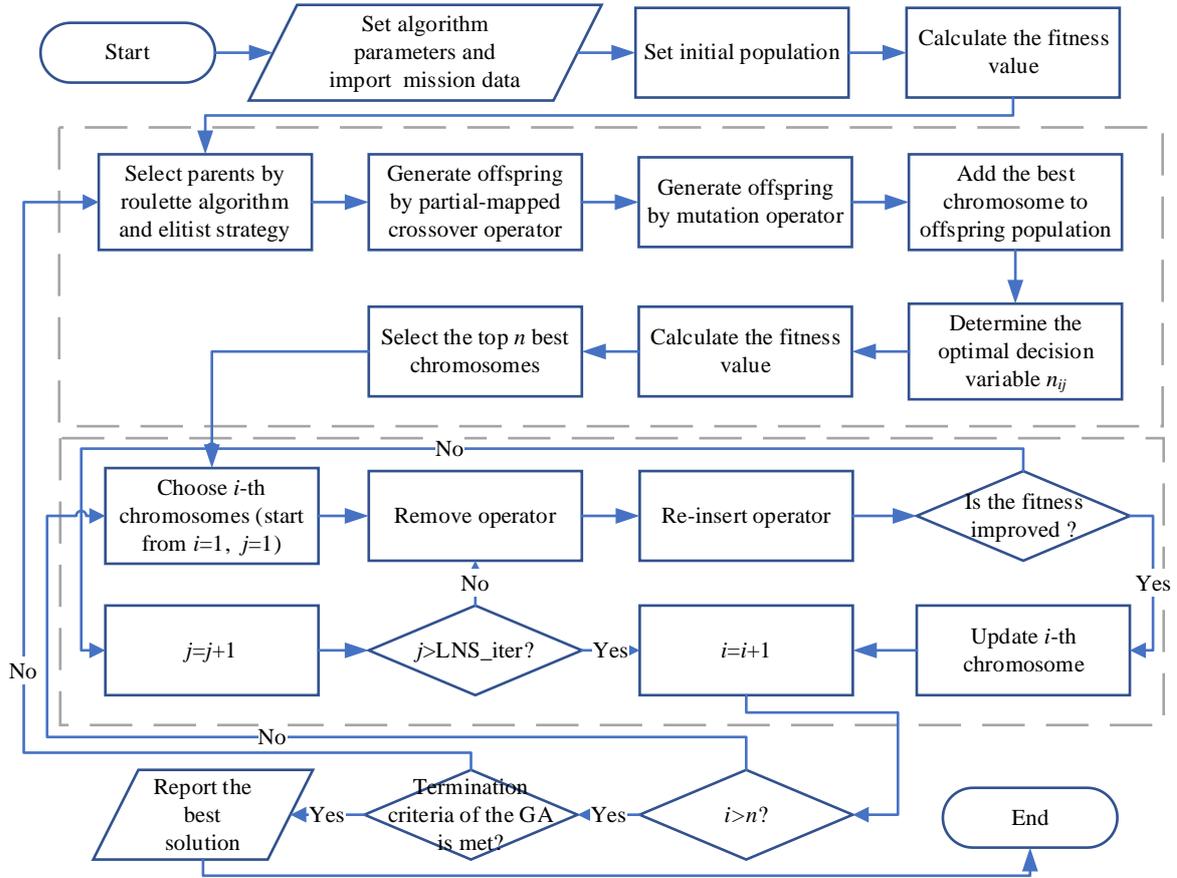

Fig. 9 Flowchart of the proposed LNS-GA algorithm

## 4  Numerical simulations and discussion

The simulation experiment is divided into two parts. The first part creates a real-world scenario by choosing 14 Chinese already launched GEO satellites as the target satellite and 2 SSc are randomly created. This scenario is used to verify the effectiveness of the proposed algorithm and determine the best parameters. The second part is used to further testing the performance of the LNS-AGA and make a comparison with other algorithms. In this part, several scenarios with different mission deadline settings and different problem scales are randomly generated. All the coded in the MALTAB 2018b environment and tested on a PC with 2.6 GHz Intel Core i5 and 8GB RAM running on the Windows 10 operating system.

## 4.1 Case study

A real-world scenario is firstly established in this part. This scenario chooses 14 already launched Chinese GEO satellites as the target satellite including the famous Beidou series, Fengyun series and Tianlian series. Those target satellites are all run on the GEO with different orbit plane and phase. The two SSc are set randomly with different inclination and true anomaly. The detailed orbit parameter is shown in Table 2. Every SSc can produce a maximum total speed increment of 1000m/s. The mission scenario starts on 12 March 2021 04:00:00 UTCG, and the true anomaly of all SSc and targets in Table 2 is exactly related to this moment. Every targets' mission deadline is set to 11 April 2021 04:00:00 UTCG, so the mission duration is exactly 30 days. The on-orbit repairing time for every target is set to 20 hours.

The parameters for LNS-AGA are set as follows: the minimum iteration number is set to 100, the population size is set to 100, the crossover probability adaptively changes from 0.7 to 0.9, the mutation probability adaptively changes from 0.01 to 0.2, the remove rate for Destroy method is 30%, the iteration number of LNS procedure is fixed to 2. The stopping criteria consist of two parts: the iteration number must exceed the minimum iteration number and the fitness value must have no change in the latest 50 iterations. Noticed that the parameter of the GA part is determined by conducting several experiments and the parameter of LNS part is determined based on Ref.[24, 26].

Besides, the parameter $n$ connecting the AGA part and LNS part in Fig. 9 which represents the gap rate between GA and LNS needs to be determined further. We can find the best parameter value with different options of $n$. Thus, we use the generated real-world to find the best parameter. Then, the computational results are obtained in 10 executions when different numbers of $n$ are tested. The results are reported in Fig. 10 with a box-plot form. From Fig. 10 we can see that when the gap rate between GA and LNS chooses 10% or 15% we can get better performance of the algorithm, by considering the computational time, we choose 10% as the parameter's best value.

Table 2 Orbit parameter of SSc and target satellite

| ID | Name | Inclination(deg) | RAAN(deg) | True Anomaly (deg) |
|---|---|---|---|---|
| -1 | SSc1 | 0 | 0 | 0 |
| -2 | SSc2 | 5 | 0 | 160 |
| 1 | Beidou2_G7 | 1.602 | 66.76 | 278.273 |
| 2 | Beidou2_G8 | 0.306 | 328.06 | 156.03 |
| 3 | Beidou_G1 | 1.801 | 45.112 | 252.161 |
| 4 | Beidou_G2 | 7.77 | 52.634 | 328.007 |
| 5 | Beidou_G3 | 1.895 | 52.106 | 274.212 |
| 6 | Beidou_G4 | 1.066 | 59.651 | 144.684 |
| 7 | Beidou_G5 | 1.455 | 67.407 | 288.524 |
| 8 | Beidou_G6 | 1.860 | 85.654 | 319.304 |
| 9 | Chinasat_11 | 0.092 | 103.257 | 331.948 |
| 10 | Fengyun_2E | 5.009 | 68.044 | 285.074 |
| 11 | Fengyun_2F | 2.806 | 83.11 | 224.488 |
| 12 | Tianlian1_01 | 4.816 | 71.744 | 337.758 |
| 13 | Tianlian1_02 | 2.211 | 74.985 | 229.245 |
| 14 | Tianlian1_03 | 0.998 | 98.186 | 230.86 |

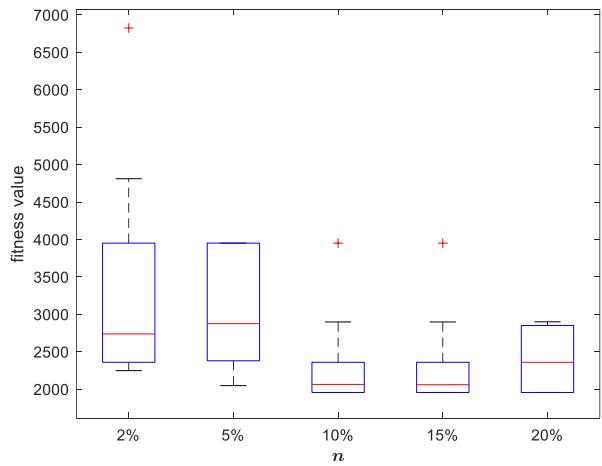

Fig. 10 The effect of gap rate between GA and LNS

After all the parameter is determined, the mission results for the real-world scenario are obtained by executing the algorithm 20 times. In addition, the problem is also solved by a general GA, which has the same parameter values and stopping criteria which LNS-AGA. The experimental results are shown in Fig. 11-Fig. 13 and Table 4.

Fig. 11 and Fig. 12 show one of the best iteration curves in 20 executions of LNS-AGA and general GA, respectively. We can that LNS-AGA converges to the optimal fitness value 1958.36 in about 50 generations while the general GA takes nearly 140 generations converges to an infeasible solution with 2285.59 fitness value. Those two figures indicate that the LNS-AGA has a faster converge rate and can get to a better solution than the general GA. Fig. 13 shows the on-orbit repairing order of the two SSc, we can see that all the targets have been repaired. All the on-orbit repair mission is achieved within the mission deadline time. Table 3 gives detailed mission results obtained by LNS-AGA, including the on-orbit repair of every SSc, the $\Delta v$ impulse vector in ECI for every orbit maneuver that can directly send to the SSc for mission execution. As a comparison, the mission planning results in Ref.[9-10] only gives a refueling route of every SSc, which needs to be detailed further for real SSc mission execution. From Table 3 we can also see that the SSc1 used total $\Delta v$ cost of 992.6817m/s and the SSc2 used total $\Delta v$ cost of 963.6836m/s, the two SSc both satisfies the maximum total velocity increment constraint. The above experimental results fully demonstrate the effectiveness of LNS-AGA in a real-world mission.

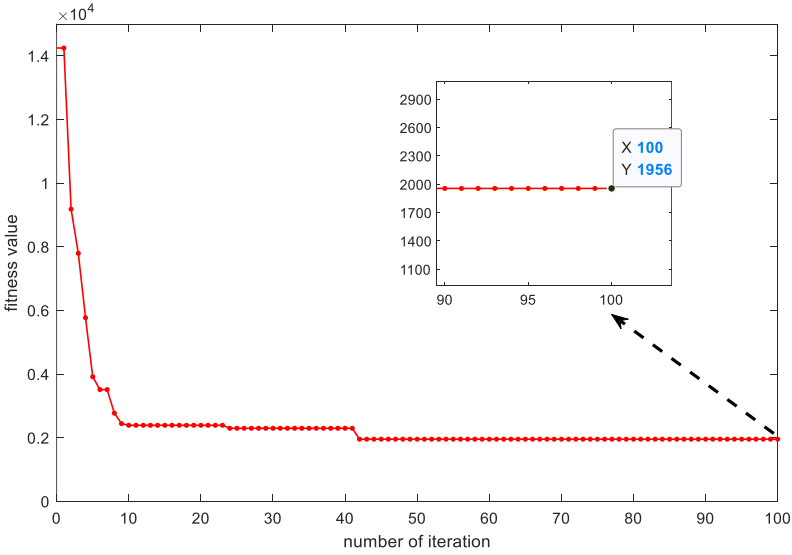

Fig. 11 Convergence curve of the LNS-AGA

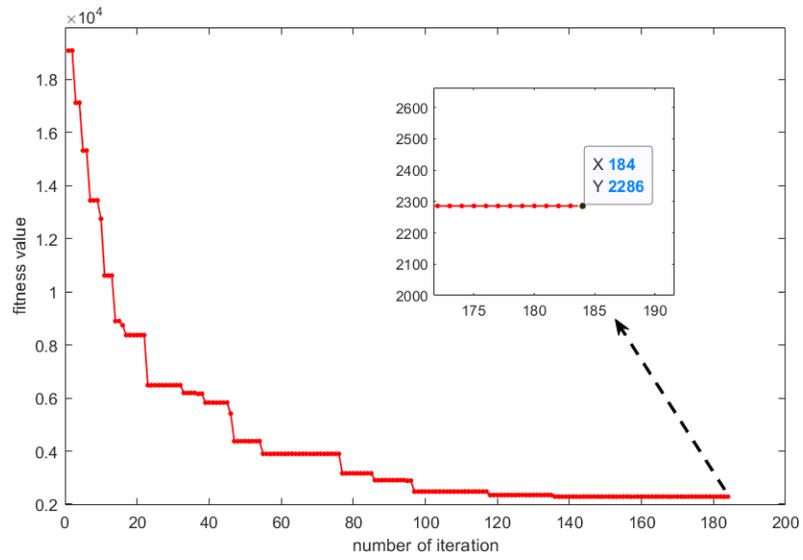

Fig. 12 Convergence curve of the general GA

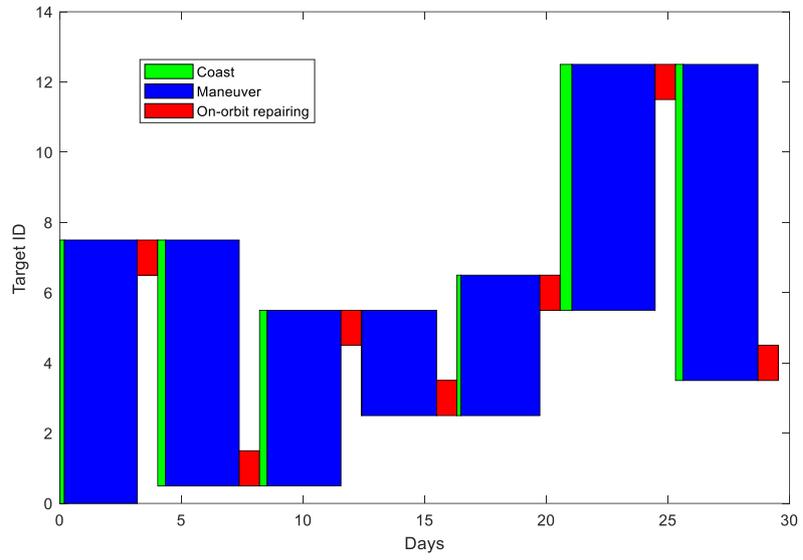

(a) SSc1 on-orbit repairing order

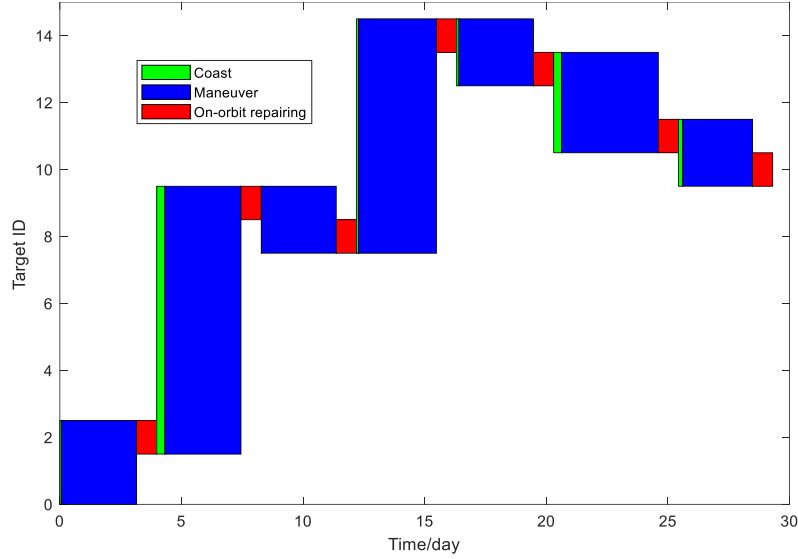

(b) SSc2 on-orbit repairing order

Fig. 13 on-orbit repairing order obtained by LNS-AGA

Table 3 Detailed mission planning results obtained by LNS-AGA

| SSc name | Target name | $\Delta v_1$ /(m/s) | $\Delta v_2$ /(m/s) | $\Delta v_1$ impulse moment | $\Delta v_2$ impulse moment | Coast time | Orbit maneuver time | $\Delta v$ cost/(m/s) |
|---|---|---|---|---|---|---|---|---|
| SSc1 | Beidou_G5 | [-2.6352, 1.0965, 78.1669] | [2.793, -1.1622, -0.07684] | 03/12 08:28:55 | 03/15 08:33:40 | 4h28m54s | 72h4m45.6102s | 81.2452 |
| | Beidou2_G7 | [-11.2548,6.4259,8.309] | [11.888,-6.7575,-0.38007] | 03/16 12:11:49 | 03/19 13:02:10 | 7h38m8s | 72h50m20.807s | 29.0745 |
| | Beidou_G3 | [-0.29144, -9.8016, -28.7957] | [-0.50225, 11.4707, 0.24622] | 03/20 16:36:53 | 03/23 17:21:43 | 7h34m42s | 72h44m50.7259s | 41.9039 |
| | Beidou_G1 | [-23.8166, -11.1914, -12.8287] | [23.492, 11.466, -0.2689] | 03/24 13:59:11 | 03/27 15:43:35 | 0h37m28s | 73h44m23.9313s | 55.4177 |
| | Beidou_G4 | [35.963, -72.9405,42.4548] | [-35.4784, 72.9763,1.2558] | 03/28 15:40:07 | 03/31 21:39:25 | 3h56m31s | 77h59m17.528s | 172.8922 |
| | Tianlian1_01 | [-113.5092, 29.062,213.329] | [133.5431, -35.4206, -11.62] | 04/02 05:21:09 | 04/05 15:27:58 | 11h41m44s | 82h6m49.022s | 382.0374 |
| | Beidou_G2 | [-4.3609, -17.1019, -194.5718] | [-15.2946, 30.9057, 4.218] | 04/06 19:08:23 | 04/09 20:51:26 | 7h40m24s | 73h43m3.7448s | 230.1108 |
| SSc2 | Beidou2_G8 | [0.76337, -44.6381, 254.0344] | [-1.184, 34.7482, 0.15414] | 03/12 05:27:36 | 03/15 07:39:23 | 1h27m35s | 74h11m47.1558s | 292.6962 |
| | Chinasat_11 | [-30.8106, -34.4337,20.2739] | [29.1489,32.5502, -0.057543] | 03/16 11:39:16 | 03/19 14:42:52 | 7h59m52s | 75h3m36.237s | 94.152 |
| | Beidou_G6 | [-26.0333, 2.439,96.0046] | [26.6314, -2.4452, -0.86837] | 03/20 10:56:15 | 03/23 12:45:23 | 0h13m22s | 73h49m8.2941s | 126.2591 |
| | Tianlian1_03 | [-63.0893, 20.2041, -47.9355] | [62.6952, -20.4725, -1.0303] | 03/24 10:45:30 | 03/27 15:37:06 | 2h0m7s | 76h51m36.2622s | 147.7308 |
| | Tianlian1_02 | [-15.8481, 10.8374,73.3511] | [16.4175,-10.2319,-0.71456] | 03/28 13:49:15 | 03/31 15:16:58 | 2h12m8s | 73h27m42.4381s | 95.1802 |
| | Fengyun_2F | [-7.3751, -1.8072, -36.8598] | [3.4451, 1.253, -0.16027] | 04/01 18:58:19 | 04/05 18:29:18 | 7h41m21.8658s | 95h30m58.7841s | 41.3032 |
| | Fengyun_2E | [-34.7182, 24.2887, -126.3111] | [25.4348, -21.055, -2.7576] | 04/06 18:37:39 | 04/09 15:20:47 | 4h8m20.8729s | 68h43m8.3194s | 166.3621 |

## 4.2 Comparation with other algorithms

For further testing the performance of the LNS-AGA, firstly, we proposed four randomly generated scenarios with different mission duration. Those scenarios consist of 10 targets and 2 SSc. The inclination of the targets and SSc are randomly generated between 0 to 10°. The RAAN and true anomaly are also generated between 0 to 360°. The maximum total speed increment constraint of every SSc increases to 2000m/s and the on-orbit repairing time for every target is set to 1 day. The mission duration is set to 10days, 15days, 25days, 35days, respectively. In addition, a general GA combined the Lambert algorithm (L-GA) which is used in Ref.[15] for calculating the orbit maneuver trajectory is also adopt for comparsion. The parameter setting of the algorithm is the same as that in 4.1. All three algorithms are tested 20 times in these four scenarios, and the experiment results are shown in Table 3.

From Table 4 we can firstly know that the LNS-AGA and the GA cannot find a feasible solution satisfying the mission deadline constraints in 10 days scenario. This is because the phasing maneuver at least need one orbital period to achieve, the total time for a single orbit rendezvous must exceed one day when the coast time is considered. For the 10 days mission duration, by subtracting the five days service time, the remaining time is certainly not enough for the five orbital maneuvers of SSc. But the L-GA can find a solution satisfying the time constraint, which is because the Lambert algorithm can find an orbit maneuver solution for any maneuver time theoretically, but usually has a tremendous cost. For other scenarios, the LNS-AGA and GA can get the same optimal solutions which is always better than L-GA. Those results also indicate that the proposed orbital maneuver strategy is better than a Lambert algorithm for the multiple spacecrafts continuous rendezvous problem while the Lambert algorithm can be applied to a stricter time constraint scenario.

Table 4 Computation results for the 3 algorithms in 4 different mission duration scenarios

| | Mission duration | 10 days | 15 days | 25 days | 35 days |
|---|---|---|---|---|---|
| GA | Minimum $\Delta v$ cost | | **2895.1896** | **2006.0872** | **1793.0918** |
| | Average $\Delta v$ cost | | 2976.8286 | 2147.6172 | 1834.1481 |
| | Standard deviation of $\Delta v$ cost | | 110.2766 | 59.745 | 49.7379 |
| | Proportion of feasible solutions | | 100% | 100% | 100% |
| LNS-AGA | Minimum $\Delta v$ cost | | **2895.1896** | **2006.0872** | **1793.0918** |
| | Average $\Delta v$ cost | | 2934.6426 | 2093.5153 | 1858.9642 |
| | Standard deviation of $\Delta v$ cost | | 62.7611 | 74.5713 | 49.5114 |
| | Proportion of feasible solutions | | 100% | 100% | 100% |
| Lambert algorithm-based GA | Minimum $\Delta v$ cost | 4070.7295 | 3248.4986 | 2534.6264 | 2114.2692 |
| | Average $\Delta v$ cost | 5590.8173 | 4186.3043 | 3136.5433 | 2656.2307 |
| | Standard deviation of $\Delta v$ cost | 1135.0188 | 572.8114 | 252.8211 | 266.125 |
| | Proportion of feasible solutions | 0% | 35% | 100% | 100% |

In order to further test the performance of the algorithm under different problem scale, based on the 15 days scenario, two larger scenarios are created including 20 targets, 4 SSc and 30 targets, 5 SSc, respectively. Other parameters for the scenarios and algorithm are the same as the first comparative experiment. The results are shown in Table 5.

In the 20 targets, 4 SSc scenario, the LNS-AGA and GA can always get a feasible solution. At the same time, the L-GA cannot find a feasible solution that satisfies the maximum $\Delta v$ cost constraint. The minimum $\Delta v$ cost obtained by LNS-AGA is about 7% lower than that obtained by GA. In the 30 targets, 5 SSc scenario, the similar conclusions can be obtained. The GA and L-GA cannot find a feasible solution but the LNS-AGA can get 9 feasible solutions for 20 times operation. Moreover, the $\Delta v$ cost obtained by LNS-AGA is obviously better than the other two algorithms. The experiment results also indicate that the LNS-AGA has a better performance in a large-scale mission scenario. The superiority of the algorithm proposed in this paper is further verified.

Table 5 Computation results for the 3 algorithms in 3 different scale scenarios

| | Problem scale | 10 targets with 2 SSc | 20 targets, 4 SSc | 30 targets, 5 SSc |
|---|---|---|---|---|
| GA | Minimum $\Delta v$ cost | **2895.1896** | 5409.8627 | 10467.7218 |
| | Average $\Delta v$ cost | 2976.8286 | 6069.7308 | 11431.7632 |
| | Standard deviation of $\Delta v$ cost | 110.2766 | 377.9312 | 583.7947 |
| | Proportion of feasible solutions | 95% | 100% | 0% |
| LNS-AGA | Minimum $\Delta v$ cost | **2895.1896** | **5056.9667** | **9107.8588** |
| | Average $\Delta v$ cost | 2934.6426 | 5337.1566 | 9591.5131 |
| | Standard deviation of $\Delta v$ cost | 62.7611 | 217.6533 | 332.3261 |
| | Proportion of feasible solutions | 100% | 100% | 45% |
| Lambert algorithm-based GA | Minimum $\Delta v$ cost | 3248.4986 | 10363.6862 | 21012.195 |
| | Average $\Delta v$ cost | 4186.3043 | 11992.1061 | 27307.0592 |
| | Standard deviation of $\Delta v$ cost | 572.8114 | 1072.7373 | 3375.246 |
| | Proportion of feasible solutions | 35% | 0% | 0% |

## 5 Conclusions

In this paper, a many-to-many on-orbit repairing mission planning problem is studied. The time-dependent mixed orbital rendezvous strategy combing planar change and phasing maneuver is designed for SSc rendezvous to the target satellite. Based on the orbital rendezvous strategy, the mission planning model is simplified to an integer programming model and be established referring to the VRPTW model. A hybrid LNS-AGA algorithm combining a large neighborhood search algorithm and adaptive genetic algorithm is then designed to solve the problem. By conducting a real-world scenario consisting of 14 lunched Chinese GEO satellites and 2 SSc, the effectiveness of the algorithm is verified. Further, several comparative experiments indicate that the proposed LNS-AGA can always get better solutions and has stronger search ability than a general genetic algorithm, and the orbital rendezvous strategy for GEO spacecraft proposed in this paper can get a more fuel-efficient orbit than the Lambert algorithm used in Ref. [15]. The optimization of the SSc's initial orbit will be studied and the maximum velocity increment constraint could be considered in the future work.

## References

[1] P. Rousso, S. Samsam, et.al., A mission architecture for On-Orbit servicing industrialization,in 2021 IEEE Aerospace Conference (50100),IEEE, 2021, pp. 1-14.

[2] J.P. Davis, J.P. Mayberry, et.al., On-orbit servicing: Inspection repair refuel upgrade and assembly of satellites in space, The Aerospace Corporation, report, (2019).

[3] W. Li, D. Cheng, et.al., On-orbit service (OOS) of spacecraft: A review of engineering developments, Prog. Aerosp. Sci., 108 (2019) 32-120.

[4] D.E. Hastings, C. Joppin, On-orbit upgrade and repair: The hubble space telescope example, J. Spacecr. Rockets, 43 (2006) 614--625.

[5] R. Qi, A. Shi, et.al., Coulomb tether double-pyramid formation, a potential configuration for geostationary satellite collocation, Aerosp. Sci. Technol., 84 (2019) 1131--1140.

[6] Y. Zhou, Y. Yan, et.al., Multi-objective planning of a multiple geostationary spacecraft refuelling mission, Eng. Optimiz., 49 (2017) 531-548.

[7] J. Zhang, G.T. Parks, et.al., Multispacecraft refueling optimization considering the j2 perturbation and window constraints, Journal of Guidance, Control, and Dynamics, 37 (2014) 111-122.

[8] X. Chen, J. Yu, Optimal mission planning of GEO on-orbit refueling in mixed strategy, Acta Astronaut., 133 (2017) 63-72.

[9] T. Zhang, Y. Yang, et.al., Optimal scheduling for location geosynchronous satellites refueling problem, Acta Astronaut., 163 (2019) 264-271.

[10] Y. Zhou, Y. Yan, et.al., Optimal scheduling of multiple geosynchronous satellites refueling based on a hybrid particle swarm optimizer, Aerosp. Sci. Technol., 47 (2015) 125-134.

[11] L. H., B. H., Optimization of multiple debris removal missions using an evolving elitist club algorithm, IEEE Transactions on Aerospace and Electronic Systems, 56 (2020) 773-784.

[12] J. Bang, J. Ahn, Multitarget rendezvous for active debris removal using multiple spacecraft, J. Spacecr. Rockets, 56 (2019) 1237-1247.

[13] H. Shen, T. Zhang, et.al., Optimization of active debris removal missions with multiple targets, J. Spacecr. Rockets, 55 (2018) 181-189.

[14] J. Yu, X. Chen, et.al., Optimal scheduling of GEO debris removing based on hybrid optimal control theory, Acta Astronaut., 93 (2014) 400-409.

[15] K. Daneshjou, A.A. Mohammadi-Dehabadi, et.al., Mission planning for on-orbit servicing through multiple servicing satellites: A new approach, Adv. Space Res., 60 (2017) 1148-1162.

[16] T.S.D. Jonchay, H. Chen, et.al., Framework for modeling and optimization of On-Orbit servicing operations under demand uncertainties, ASCEND, 2021.

[17] L. Federici, A. Zavoli, et.al., Evolutionary optimization of multirendezvous impulsive trajectories, Int J. Aerospace Eng, 2021 (2021) 1-19.

[18] G. He, R.G. Melton, Multiple small-satellite salvage mission sequence planning for debris mitigation, Advances in the Astronautical Sciences, 171 (2019) 4229--4244.

[19] B. Ombuki, B.J. Ross, et.al., Multi-objective genetic algorithms for vehicle routing problem with time windows, Appl. Intell., 24 (2006) 17-30.

[20] H. Chen, K. Ho, Integrated space logistics mission planning and spacecraft design with mixed-integer nonlinear programming, J. Spacecr. Rockets, 55 (2018) 365-381.

[21] Z. Zheng, J. Guo, et.al., Distributed onboard mission planning for multi-satellite systems, Aerosp. Sci. Technol., 89 (2019) 111-122.

[22] S. Wang, L. Zhao, et.al., Task scheduling and attitude planning for agile earth observation satellite with intensive tasks, Aerosp. Sci. Technol., 90 (2019) 23-33.

[23] A. Baniamerian, M. Bashiri, et.al., Modified variable neighborhood search and genetic algorithm for profitable heterogeneous vehicle routing problem with cross-docking, Appl. Soft Comput., 75 (2019) 441-460.

[24] S. Ropke, D. Pisinger, An adaptive large neighborhood search heuristic for the pickup and delivery problem with time windows, Transport. Sci., 40 (2006).

[25] D. Pisinger, S. Ropke, Large neighborhood search, Handbook of metaheuristics, Springer2010. pp. 399--419.

[26] P. Shaw, Using constraint programming and local search methods to solve vehicle routing problems, Springer Berlin Heidelberg, Berlin, Heidelberg, 1999. pp. 417-431.

[27] S. Hu-Li, H. Yan-Ben, et.al., Beyond life-cycle utilization of geostationary communication satellites in end-of-life, Satellite Communications, Nazzareno Diodato (Ed), Intech, (2010) 323--365.

[28] L.J. Friesen, A.A. Jackson IV, et.al., Analysis of orbital perturbations acting on objects in orbits near geosynchronous Earth orbit, Journal of Geophysical Research: Planets, 97 (1992) 3845--3863.

[29] M. Mitchell, An introduction to genetic algorithms, MIT press1998.

[30] J.H. Holland, Adaptation in natural and artificial systems: An introductory analysis with applications to biology, control, and artificial intelligence, MIT press1992.

[31] D.E. Goldberg, R. Lingle, Alleles, loci, and the traveling salesman problem, Carnegie-Mellon University Pittsburgh, PA, 1985, pp. 154--159.